# Developing a method for elaboration the scenarios related with sustainable products lifecycle

Hayder Alhomsi, Peggy Zwolinski,
G-SCOP Laboratory, 46 av Félix Viallet – 38031 Grenoble Cedex, France
Hayder.Alhomsi@g-scop.inpg.fr_+33688316909
Peggy.Zwolinski@g-scop.inpg.fr_+33673931381

## Abstract
This article aims at presenting our objective that is to use DfD rules earlier during the design process. Indeed, during the conceptual design phase, designers don't have simple qualitative tools or methods to evaluate their products. There are guidelines that are very useful in a first approach to give some objectives, but there is no quantitative indicators associated to these rules to consider the disassembly aspects when the first choices are realised for the product.
So we will present that to use DfD rules during the conceptual design phase, we first have:
- to identify which kind of rules can be applied when designers only have a functional representation of their product.
- to create the necessary indicators to evaluate these rules depending on designers choices.

We think that this approach is usable for many DfX rules either if we only consider in this paper DfD rules.

## Introduction

Design for disassembly (DFD) is an important design concept to make products more effective for maintenance, recycling, remanufacturing and other related processes [1, 2]. At the same time it is friendly for the environment and its related specifications because it enables to improve the product End-of-Life recovery of parts and materials. One of the important issues in DFD is related to the selection of the characteristics of components and relations between them.

The most comprehensive work on Design for Disassembly has been carried out by Boothroyd, Wittenberg G, Beitz [2] [3] and VDI who have identified the more detailed areas associated with Design for Recycling:

- Designing for ease of disassembly (to enable the removal of parts without damage).
- Designing for ease of purifying (to ensure that the purifying process does not damage the environment).
- Designing for ease of testing and classifying (to make it clear as to the condition of parts which can be reused and to enable easy classification of parts through proper markings).
- Designing for ease of reconditioning (this supports the reprocessing of parts by providing additional material as well as gripping and adjusting features).

## DfD Approach

### Existing approach for DfD consideration into the design process

The design process starts from the product specification as the result of client's needs and society statement. During the conceptual design phase designers try to identify the necessary elements to realise these specifications. Finally, during the detailed design phase, designers have to define precisely the product taking into account all its life cycle processes.
For the analysis of the product from a disassembly point of view, designers have to consider product's EoL scenario[4].

These scenarios are generally not considered during the Conceptual Design but during the detailed design phase [3, 5, 6]. That means that if the designed product is not adapted to these scenarios designers have to repeat the conceptual design phase to do the proper modifications. Our objective in this research is to propose a method usable earlier during the design process, to help designers to integrate product end of life scenario for an integrated design.

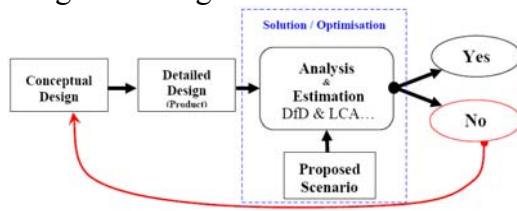

*Figure 1: Existing approach for DfD consideration during the design process*

## Proposed approach for DfD consideration into the design process

To reduce the level of non acceptable solutions at the end of the detailed design phase, we have to propose a method able to define both the product life cycle and the product structures [7]. Indeed, during the conceptual design, a functional analysis is realised and provides: (1) The main characteristics of the main components in the future product. (2) The relations between them. So, if designers define each component's LC scenario, it is possible to define a first compromise for the structure of the whole products while using DfD rules.

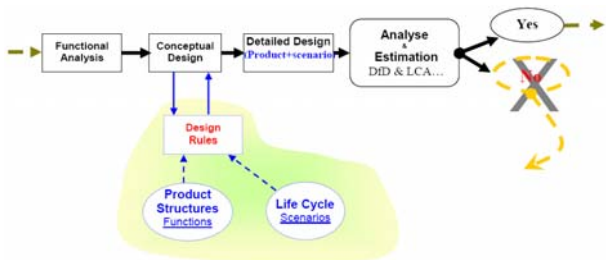

*Figure 2: Proposed approach for DfD consideration during the design process.*

## A method to use DfD rules during the conceptual design phase

Numerous rules exist to evaluate the disassembly of a product. These rules can be classified into three groups[7]:
- A first group related to the design of the relations between components. For example in this group, there is the following rule: "The relation between components should be easy to disconnect"
- A second group is related to the design of the structure of the products that is constituted of relations and components. I.e.: "Minimising the variety of component materials in the whole product", "Minimising the number of type of components in EoL".
- A third group is related with the pollution characteristic. I.e.: "The polluting components should be the first components to be disassembled", "The polluting component(s) should be disassembled from the main product by using few disassembly tools", "If there is more than one polluting component, it is preferable to disassemble them in one direction."

To improve the use of these rules during the design process we have:
- Identify the product model that can support these rules during the conceptual design phase.
- Define factors related to the design rules and weighting factors related to the end of life scenarios to evaluate the preliminary solutions from a disassembly point of view.

## The product model used during the conceptual design phase

The product model has been chosen regarding the simplest combination needed to obtain a Structure able to be disassembled. The simplest structure consists of two components and one relation (figure 3). It is the (C, R, S) model for Component, Relations between the components and Structure.

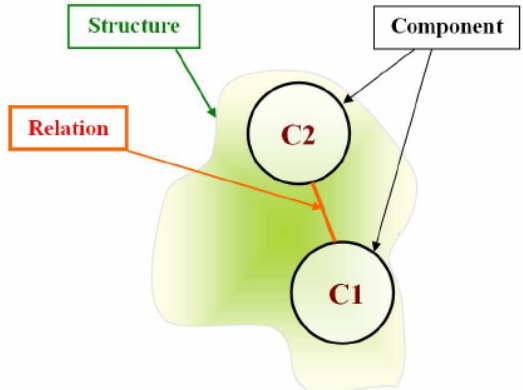

*Figure 3: The simplest product combination needed to be disassembled.*

## Indicators

To evaluate the design choices, we have decided to consider a triple indicator (Kc, Kr, Ks) related to the (C, R, S) model. For each of these three indicators, a value is assigned, related to specific characteristic. Some of these indicators have numerical values (weight, number…) and some of them are described by a literal formulation (material, EoL…).

We have defined a symbolisation for these indicators (figure 4). They are presented with the letter (K), and their group is specified with the second letter (C, R or S). The third letter is a (P) if the indicators have specific relations with polluting components. At the end of the indicator appears the abbreviation for the characteristic considered, in small letters.

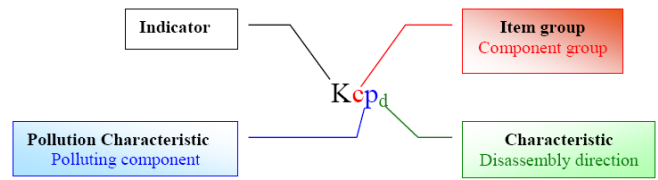

*Figure 4: Indicators' symbolisation*

For example, we have:
- $Kc_{da}$ that belongs to the component characteristics groups and gives information on the component disassembly axis.
- $Kr_{type}$ that belongs to the relations characteristics group and give information on the Type of relations. There are general types and the designer chooses the relation type from the list presented table 1.

| Relation Type | |
|---|---|
| 1 | Pull |
| 2 | Press |
| 3 | Snapping |
| 4 | Bolt & Nut |
| 5 | Thread |
| 6 | Crimp |
| 7 | Rivet |
| 8 | Sticking |

*Table 1: Types of relations and joints*

- $Ks_{Cn}$ that belongs to the structure characteristics group and give information on the total number of components in the product.

All the indicators are listed in the table 2.

## Factors

The factor is an evaluated item that is linked to a DfD rule. The factor is evaluated by a formula that uses the indicators presented in the last section and is valued as a real number belonging to an interval [0-1].

The symbolisation system of factors looks like the indicator's symbolisation system (figure 5). The factors are represented by a letter (F) and their group is specified with the second letter (R or S). The third letter is a (P) if the indicators have specific relations with polluting components. At the end of the indicator appears the abbreviation for the rule considered, in small letters

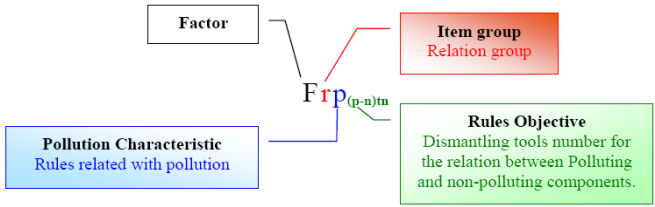

*Figure 5: Factor symbolisation*

# Proposed rules

For example, we will illustrate some of these rules:

## Rule No.1

**(THE RELATION TYPE BETWEEN COMPONENTS SHOULD BE EASY TO BE DISCONNECTED [8].)**

Evaluating this rule is possible by the following factor:

**$Fr_e$** : (Easiness factor of relation disconnection).

This factor has relationship with ($Kr_{eas}$) Disconnecting easiness degree indicator (how is it easy to break the relation between components).

This factor is estimated with the following equation:

$$Fr_e = interval\,[0 - 1] = \sum_{i=n}^{i=1}(Kr_{eas(i)})/\,n$$

n: number of relations

This factor defines the way of disconnecting the relation and how the disassembly process is easy (with taking in account the quality of the disassembled components). Manual process is preferable and it is taken as the easiest disconnecting way, reversible relation is the second way but the destructive ways (disassemble the relation or destructive the components) have the lowest values.

## Rule No.2

**(MINIMISING THE VARIETY OF COMPONENT MATERIALS IN THE WHOLE PRODUCT" [8].)**

Evaluating this rule is possible by the following factor:

**$Fs_{mws}$** : (Product similarity factor of material weight[9]).

This factor is related with each components material type ($Kc_{Mat}$ : Component material Indicator) and their weights ($Kc_W$: Component weight indicator). Each type of materials has its separated factor. The value used in this factor is the material biggest weight (the largest quantity of any material). This factor is estimated with the following equation:

$$Fs_{mws} = Internal\,(0\text{-}1) = (Ks_{mws})\,/\,(Ks_{Wtot});$$

$$\sum_{i=n}^{i=1} Fs_{mws(i)} = 1$$

$$Fs_{mws(1)} + Fs_{mws(2)} + Fs_{mws(3)} + \ldots = 1$$

$Ks_{mws}$ : Product material weight similarity Indicator.
$Ks_{Wtot}$ : Product total weight Indicator.
$Fs_{mws}(1)$ : Percentage of material weight (M1) in the product (S).
$Fs_{mws}(2)$ : Percentage of material weight (M2) in the product (S).
$Fs_{mws}(3)$ : Percentage of material weight (M3) in the product (S).
…

## Rule No.2

**(THE POLLUTING COMPONENTS SHOULD BE THE FIRST DISASSEMBLED COMPONENTS TO BE DISASSEMBLED [10].)**

Evaluating this rule is possible by the following factor:

**$Fsp_{(p\text{-}ns)r}$:** (Relation Factor between polluting component and non-polluting components[7]).

This factor is related with (NP-P) relation type (relations between polluting component and non-polluting components). The factor has inverse relationship with the number of indicators (Krp) (*Indicator Relation-Pollution condition*) which have value (NP-P).

The factor is related neither with the number of internal-relation between the polluting components (P-P) nor with the number internal-relation between non-polluting components (NP-NP). This factor helps the designer evaluating the disassembly case of polluting components as an external part. This factor is estimated with the following equation:

$$Fsp_{(p-ns)r} = \text{interval } [0\text{-}1] = \frac{1}{\sum (Kr_p)_{;(NP-P)}}$$

## Factor total and Weighting indicators

After identifying the rules by factors and obtaining the value of each factor, we have to calculate the main factor (FACTOR TOTAL). This factor total is specific for the whole product and represents the aggregation in one value of all the factors.

In this article, the suggested rules addressed two concepts:
- Rules to adapt the DfD principles (Design for Disassembly principles).
- Rules to ameliorate the Product EoL (from de-pollution and recycling points of view)

The final factor value (value of factor total) represents the satisfaction of designer to adapt these two points of view in the product design. This value refers also to the disassembly ability of the proposed design. This ability takes into account the end of life (EoL) and the amelioration of the life cycle (LC) scenario in the same time.

The factor total value can be evaluated by giving each factor a different weighting value (the value of the weighting indicator related with designer point of view and by dividing the total with the sum of the weightings.

$$Fs_{tot} = \text{interval } [0\text{-}1] = \frac{\sum_{i=n}^{i=1} I_i * F_i}{\sum_{i=n}^{i=1} I_i}$$

**$Fs_{tot}$**: Factor total for the whole product (S)
**$I_i$** : Weighting indicator for the factor (Fi)
**$F_i$** : Factor of one realised rule
**n** : Number of used factors

For giving a proper justification for the values of factor, it is very important to link these factors with obvious criteria: links with known database (standards, limits and reference indicators), reference to the specialty of each product and the experience of the designer (designers and researches related with new product, processes or materials). The value of the weighting indicators is defined by the designer itself. He chooses the value according to the customer's needs and the designing specification.

In this research the scale of weighting indicator will be assumed as a scale of ten ( X /10); the most important will take 10 and less important will take 0 depending on the designer point of view.

$$I_i = \text{Value (real numeral)}$$

The factor is an evaluated item that is linked to a DfD rule. The factor is evaluated by a formula that uses the indicators presented in the last section and is valued as a real number belonging to an interval [0-1].

The symbolisation system of weighting indicator looks like the indicator's symbolisation system (figure 6). The weighting indicators are represented by a letter (I) and their group is specified with the second letter (R, C or S). The third letter is a (P) if the indicators have specific relations with polluting components. At the end of the indicator appears the abbreviation for the main specific characteristic of the weighting indicator considered, in small letters

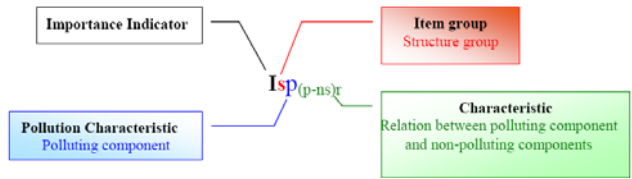

*Figure 6: Weighting indicator symbolisation system.*

The next figure (figure 7) demonstrates the factor value (Fstot) as a percentage (as mentioned before) to be compared with the proper limits of the design satisfaction. The value near (0) appears when the proposed design is bad, and the value near 100% appears when the design is good.

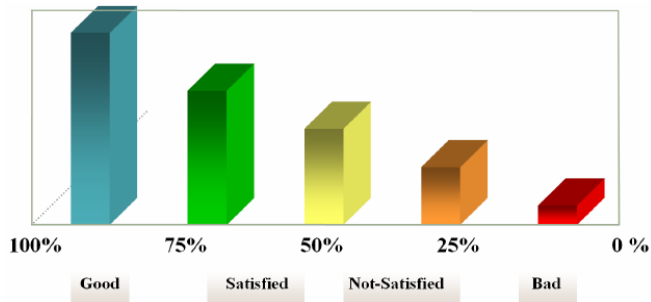

*Figure 7: Factor total limits for general satisfaction.*

The factor total is called (DISASSEMBILITY FACTOR) [14]. The objective from factor formulation is to provide the designer first estimation for his design. This estimation is related to how much the design adapts the aspects of environment and DfD principle during the conceptual design phase. We will see an application of this evaluation in the next section.

## Case study

In this case study, we will consider the design of plastic potable water. The main components of this bottle have been defined during the conceptual design phase as in the Table 3. During the design process, a functional block diagram has been establish (figure 8) and shows the functional components and their relations that are necessary to know to define the (R,C,S) characteristics.

| Component | | material | |
|---|---|---|---|
| 1 | Main body | Polyethylene terephthalate | (PET) |
| 2 | Cap | Polypropylene | (PP) |
| 3 | Cap's ring | Polypropylene | (PP) |
| 4 | Fixation ring | Stainless- Steel | (St) |
| 5 | Adhesive label | Polyvinyl chloride | (PVC) |

*Table 3: Components and materials of potable water bottle*

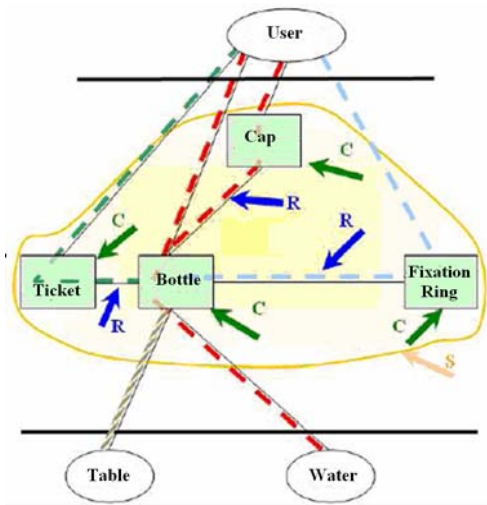

*Figure 8: Plastic potable water bottle BDF-S and the triple (R,C,S)*

### Analysis

In a first analysis, designers have compared two solutions: the solution described in the table 3 and a solution with a material modification for the fixation ring that becomes realised with plastic (PP). For the two cases, the EoL is reuse.

In a second analysis designers have modified the whole EoL scenario and chosen the recycling. A new evaluation of the weighting factors has been done.

## Results for the first Analysis

In this first analysis, only factors have changed according to DfD rules.

The weighting indicators have been defined (table 4) as well as the factors and the final result for the initial product has been calculated: Ftot = 55.80%.

For the second calculation, three factors have been modified (Table 5). They are all related with the materials and weights characteristics. The factor total for the whole product has been evaluated: Ftot = 69.61%.

## Results for the second analysis

In this case only the weighting indicators have changed (table 6). The material of the fixation ring is still Stainless-Steel. The factor total for the whole product has been evaluated: Ftot = 48.99%.

For the second calculation the material of the fixation ring is Polypropylene (PP). The factor total for the whole product has been evaluated: Ftot = 68.16%.

## Conclusion (case study)

The two analyses show the influence of changing the structural design and the LC scenario. Depending on the chosen product, it is the material change that induces the most **important** modification on the total factor (figure 9). This can be explained because of the small number of parts in this example, but in most of the cases, the two types of modification should be very interlinked.

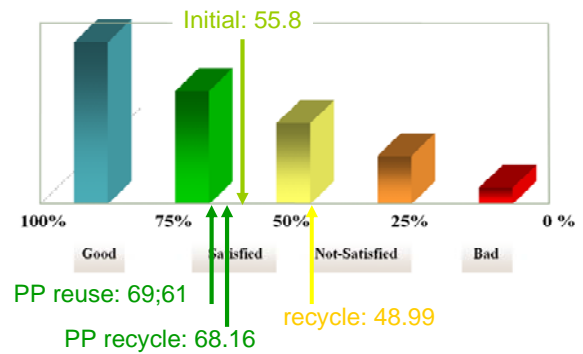

*Figure 9: Results of the factor total calculation for the different analysis*

## Conclusion

Our method is based on the establishment of a factor called "disassembly factor" that can be used during the conceptual design phase to help designers to better integrate the disassembly during the design process.

We have proposed to calculate this factor while using:
- DfD rules that are usually used later in the design process
- a (C,R,S) model for the product that supports the necessary data for the disassembly factor calculation
- Indicators and factors that have to be considered to evaluate the disassembly.
- weighting factors to weight the proposition depending on the product life cycle

The factor total gives to the designer a primer product estimation value from a disassembly point of view. It is a mean to avoid numerous trials/errors sequences when designing the structure of the product.

## References

1. Zwolinski, P.H., N. Brissaud, D., *How to integrate End of Life Disassembly Contraintes in Early design stage?* Int. Sem. CIRP LCE Belgrad, 2004.
2. HAOUES, N., *Contribution à l'intégration des contraintes de désassemblage et de recyclage dés la première phase de conception de produits. ."* (2006).
3. Beitz, G.P.a.W., *Engineering Design (A Systematic Approach).* 3rd edition 2007: p. 39.
4. Peggy ZWOLINSKI, G.P., Daniel BRISSUAD, *Environment and design* G-SCOP lab. Grenoble France, 2004 p. Page -223-240.

5. Jingyan Zuo, S.D., „ *An Integrated Design Environment for Early Stage Conceptual Design.*
6. F. GIUDICE, G.L.R.A.R., *Product Design for the Environmental life cycle approach.* **chapter3 (life cycle design and management)**: p. page 61-82.
7. ALHOMSI, H.Z., P., *Developing a method for elaboration the scenarios related with sustainable products lifecycle.* 2007.
8. Bhootra., A., *A Disassembly Optimization Problem.* . 2002.
9. ECODESIGN-PILOT, *Preferably use single material components and/or reduce number of different types of material*   Vienna University of Technology.
10. Dimitris Kiritsis, A.B., Paul Xirouchakis. , *Multi-criteria decision aid for product end of life options selection.* . 2003: p. page 49.

**ALHOMSI Hayder**
**Supervising: ZWOLINSKI Peggy**
**INPG, G-SCOP Laboratory**
**Grenoble –France**
**2008**

# Annexe

Table 2: Indicators group and their symbols

| Indicator | Symbol | Indicator's value | Group |
|---|---|---|---|
| Disassembly axis indicator | $Kc_{da}$ | List (X, Y, Z) | Group (C) |
| Indicator of Direction of disassembly axis | $Kc_{da}$ | Function (Positive/Negative) | Group (C) |
| Pollution condition indicator | $Kc_p$ | Function (Yes / No) | Group (C) |
| Component weight Indicator | $Kc_W$ | Value (real numeral) | Group (C) |
| Component material Indicator | $Kc_{Mat}$ | Symbol (displayed text) | Group (C) |
| Component end of life indicator | $Kc_{EoL}$ | Symbol (displayed text) | Group (C) |
| Recyclability Indicator). | $Kc_{Rcycl}$ | Function (Yes / No) | Group (C) |
| Disassembly Direction indicator of polluting components | $Kc_{pd}$ | Kcpd = Pairs (Axis, direction) | Group (C) |
| Disconnecting easiness degree indicator | $Kr_{eas}$ | List (x1, x2, x3, , x4) | Group (R) |
| Relation type indicator | $Kr_{type}$ | List (x1, x2, x3…) | Group (R) |
| Indicator Relation-Pollution condition | $Kr_p$ | List ( P-P, NP-P & NP-NP) | Group (R) |
| Tools number indicator for disconnecting component relation | $Kr_{tn}$ | Value (real numeral) | Group (R) |
| Indicator of components number | $Ks_{Cn}$ | Value (1, 2, 3…) | Group (S) |
| Indicator of relations number | $Ks_{Rn}$ | Value (1, 2, 3…) | Group (S) |
| Indicator of polluting components number. | $Ks_{pcn}$ | Value (0, 1, 2, 3…) | Group (S) |
| Product total weight Indicator | $Ks_{Wtot}$ | Σ KcW = Value (real numeral) | Group (S) |
| Product material weight similarity Indicator | $Ks_{mws}$ | Value (real numeral) | Group (S) |
| Product material number similarity Indicator | $Ks_{mns}$ | Value (real numeral) | Group (S) |
| Weight similarity indicator of components' EoL | $Ks_{EoLws}$ | Value (real numeral) | Group (S.) |
| Product's EoL number similarity Indicator | $Ks_{EoLns}$ | Value (real numeral) | Group (S) |
| Indicator of all polluting components number | $Ks_{pcn}$ | Value (number) | Group (S) |

Table 4: Results for the initial case in the first analysis (Weighting indicators and Factors)

| | Weighting indicators | | |
|---|---|---|---|
| No, | Name | Symbol | Value |
| 1 | Weight for: Easiness for relation disconnection | $Ir_e$ | 9 |
| 2 | Weight for: Product similarity of material weight | $Is_{mws}$ | 5 |
| 3 | Weight for: Product similarity of material type | $Is_{mns}$ | 2 |
| 4 | Weight for: Product similarity for components' EoL (Weight) | $Is_{EoLws}$ | 9 |
| 5 | Weight for: Product similarity for components' EoL (type) | $Is_{EoLns}$ | 5 |
| 6 | Weight for: Relation between polluting component and non-polluting component | $Isp_{(p-ns)r}$ | 0 |
| 7 | Weight for: Dismantling tools number to disassemble polluting component | $Irp_{(p-n)tn}$ | 0 |
| 8 | Weight for: Similarity direction factor for Polluting components disassembly. | $Isp_{sd}$ | 0 |

| | Factors | | |
|---|---|---|---|
| No, | Name | Symbol | Value |
| 1 | Easiness for relation disconnection | $Fr_e$ | 58,33% |
| 2 | Product similarity of material weight | $Fs_{mws}$ | 51,72% |
| 3 | Product similarity of material type | $Fs_{mns}$ | 25,00% |
| 4 | Product similarity for components' EoL (Weight) | $Fs_{EoLws}$ | 51,72% |
| 5 | Product similarity for components' EoL (type) | $Fs_{EoLns}$ | 75,00% |
| 6 | Relation between polluting component and non-polluting components | $Fsp_{(p-ns)r}$ | 100,00% |
| 7 | Dismantling tools number to disassemble polluting component | $Frp_{(p-n)tn}$ | 100,00% |
| 8 | Similarity direction factor for Polluting components disassembly. | $Fsp_{sd}$ | 100,00% |

Table 5: Factors changes depending on the material modifications

| | Factors | | |
|---|---|---|---|
| No, | Name | Symbol | Value |
| 1 | Easiness for relation disconnection | $Fr_e$ | 58,33% |
| 2 | Product similarity of material weight | $Fs_{mws}$ | 58,82% |
| 3 | Product similarity of material type | $Fs_{mns}$ | 50,00% |
| 4 | Product similarity for components' EoL (Weight) | $Fs_{EoLws}$ | 88,24% |
| 5 | Product similarity for components' EoL (type) | $Fs_{EoLns}$ | 75,00% |
| 6 | Relation between polluting component and non-polluting components | $Fsp_{(p-ns)r}$ | 100,00% |
| 7 | Dismantling tools number to disassemble polluting component | $Frp_{(p-n)tn}$ | 100,00% |
| 8 | Similarity direction factor for Polluting components disassembly. | $Fsp_{sd}$ | 100,00% |

Table 6: Weighting indicators changes depending on the life cycle option

| | Weighting indicators | | |
|---|---|---|---|
| No, | Name | Symbol | Value |
| 1 | Weight for: Easiness for relation disconnection | $Ir_e$ | 2 |
| 2 | Weight for: Product similarity of material weight | $Is_{mws}$ | 9 |
| 3 | Weight for: Product similarity of material type | $Is_{mns}$ | 5 |
| 4 | Weight for: Product similarity for components' EoL (Weight) | $Is_{EoLws}$ | 9 |
| 5 | Weight for: Product similarity for components' EoL (type) | $Is_{EoLns}$ | 2 |
| 6 | Weight for: Relation between polluting component and non-polluting component | $Isp_{(p-ns)r}$ | 0 |
| 7 | Weight for: Dismantling tools number to disassemble polluting component | $Irp_{(p-n)tn}$ | 0 |
| 8 | Weight for: Similarity direction factor for Polluting components disassembly. | $Isp_{sd}$ | 0 |